\documentstyle[12pt,aaspp4]{article}
\begin{document}

\title{Two Confirmed Cataclysmic Variables in the Old Stellar
Cluster NGC 6791\footnote{Based
on the observations collected at the Michigan-Dartmouth-MIT (MDM)
1.3-meter telescope, the F. L. Whipple Observatory (FLWO)
1.2-meter telescope and the Multiple Mirror Telescope (MMT).}}

\author{J. Kaluzny}
\affil{Warsaw University Observatory, Al. Ujazdowskie 4,
00--478 Warszawa, Poland} 
\affil{\tt e-mail: jka@sirius.astrouw.edu.pl} 
\author{K. Z. Stanek\altaffilmark{2}, P. M. Garnavich, P. Challis} 
\affil{\tt e-mail: kstanek@cfa.harvard.edu, pgarnavich@cfa.harvard.edu,
challis@cfa.harvard.edu} 
\affil{Harvard-Smithsonian Center for Astrophysics, 60 Garden St., MS20, 
Cambridge, MA~02138} 
\altaffiltext{2}{On leave from N.~Copernicus Astronomical Center, 
Bartycka 18, Warszawa 00--716, Poland} 

\begin{abstract}

Based on new CCD photometry and spectroscopy we confirm the presence
of two cataclysmic variables (CVs) in the very old open cluster
NGC~6791. One of these variables, known as B8, was observed by
Rucinski, Kaluzny \& Hilditch (1996) to undergo a large magnitude
outburst in 1995. The spectrum of this star outside the outburst,
obtained by us with MMT, shows clearly the emission lines
characteristic for dwarf novae. We observed the second star, known as
B7, to undergo large ($\sim3\;mag$) drop in its brightness over $\sim
10\;days$.  The spectrum of B7 obtained in the high state resembles
spectra of nova-like CVs. This star a likely member of UX~UMa subtype
of CVs.  Variables B7 and B8 represent only the second and the third
cataclysmic variables known in open clusters.  Variable B7 has
observational characteristics which would make it difficult to
identify as a CV with some of the methods being currently used in
surveys for CVs in globular clusters.

\end{abstract}


\section{INTRODUCTION}

NGC~6791 is currently considered to be the one of oldest known open
clusters in the Galaxy.  Recent photometric work includes studies by
Kaluzny \& Udalski (1992, hereafter: KU), Montgomery, Janes \& Phelps
(1994), Garnavich et al.~(1994) and Kaluzny \& Rucinski (1995,
hereafter: KR95). There is a spectroscopic as well as a photometric
evidence that the cluster metallicity is a factor of 2--3 higher than
the solar value (Friel \& Janes 1993; Garnavich et al.~1994; KR95).
NGC~6791 is also one of the most massive open clusters
currently known with the total mass of observed stars exceeding $4000
M_\odot$ (KU).

NGC~6791 is unique among old open clusters in harboring a group of hot
subdwarfs. KU identified 8 faint blue stars in the cluster field
based on $BVI$ photometry. They suggested that these objects are hot
subdwarfs belonging to the cluster. Two additional candidates for hot
subdwarfs were identified by KR95 who provided also $U-B$ colors for
all candidates. Some of these blue stars were observed by Liebert et
al.~1994 and indeed found to be hot subdwarfs and likely members of
the cluster (see also Green et al.~1996).

Most of the previous studies concentrated on the age and metallicity 
of the cluster and only two searches for variable stars in NGC~6791 have
so far been conducted (Kaluzny \& Rucinski 1993, hereafter: KR93;
Rucinski, Kaluzny \& Hilditch 1996, hereafter: RKH), leading to
discovery of 8 contact binaries and 14 other variables, with many
among the latter being detached eclipsing systems. A fraction of these 
variables are likely to be foreground field stars.

The main motivation of the present study was to provide a continuation
to variability studies of KR93 and RKH. Our intention was to monitor
long-term variability of stars in NGC~6791. The detailed results of the
CCD photometry obtained for the $10.9\times10.9\;arcmin$ region of the
cluster will be presented elsewhere (Stanek \& Kaluzny 1997, in
preparation). In this Letter we confirm the presence of {\em two}
cataclysmic variables in NGC~6791, which presence was suspected based
on earlier data (KR93; RKH; Liebert et al.~1994). Although several
cataclysmic variables are either detected or suspected in globular
clusters (Grindlay et al.~1995; Livio 1996;  Bailyn et al.~1996), we
note that these two stars are the second and third cataclysmic
variables known in old open clusters. The first such system was
detected by Gilliland et al.~(1991) in M~67.

\section{NEW OBSERVATIONS}

We obtained photometric data on 30 nights between September~8 and
November~1, 1996.  NGC~6791 was primarily observed with the
McGraw-Hill 1.3-meter telescope at the MDM Observatory. We used the
front-illuminated, Loral $2048^2$ CCD Wilbur (Metzger, Tonry \&
Luppino 1993), which at the $f/7.5$ station of the 1.3-meter has a
pixel scale of $0.32\;arcsec/pixel$ and field of view of roughly
$10.9\;arcmin$. We used Kitt Peak Johnson-Cousins $BVI$ filters.  Some
data for NGC~6791 were also obtained with the 1.2-meter telescope at
the FLWO, where we used ``AndyCam'' with thinned, back-side
illuminated, AR coated Loral $2048^2$ CCD (Caldwell et al.~1996).  The
pixel scale happens to be essentially the same as at the MDM 1.3-meter
telescope. We used standard Johnson-Cousins $BVI$ filters (Caldwell et
al.~1996).  We obtained for NGC~6791 useful data during 25 nights at
the MDM, collecting a total of 79 $450\;sec$ exposures in $V$ and 50
$300\;sec$ exposures in $I$. We obtained additional data during 5
nights at the FLWO, collecting a total of 30 $450\;sec$ exposures in
$V$ and 4 $300\;sec$ exposures in $I$.

The profile photometry was extracted using Daophot/Allstar programs
(Stetson 1987). More detailed description of applied procedures is
given in Kaluzny et al.~(1997).  Transformations to the standard
$VI_{C}$ system were obtained by calibrating out the color terms using
Landolt (1992) fields and determining the zero-point offsets with the
KU data.

Spectra of B7 and B8 were obtained with the Multi-Mirror Telescope
(MMT) and Blue Channel Spectrograph. Exposures of B7 were taken on
April~8, May~2 and 3, 1997, while data for B8 were obtained on
April~7, 1997. A 300 line/mm grating was used with a $1\;arcsec$ slit
at the parallactic angle providing a resolution of 7\AA\ FWHM and
spectral coverage from 3200\AA\ to 8000\AA . The seeing was better
than $1.0\;arcsec$ on all the nights, so the slit losses were small.

\section{RESULTS FOR THE VARIABLES}

\subsection{The cataclysmic variable B8}

\begin{figure}[t]
\plotfiddle{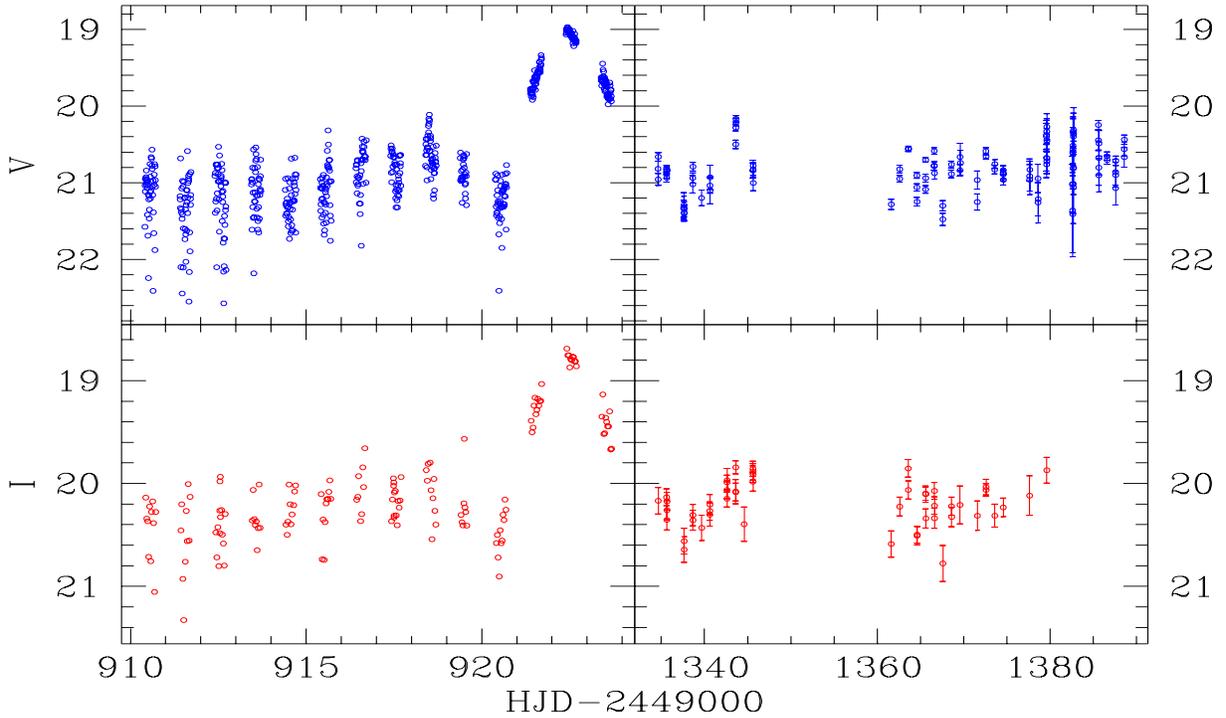}{8cm}{-90}{60}{50}{-240}{285}
\caption{$V$ (upper panels) and $I$ (lower panels) light curves of the
cataclysmic variable B8. Note that the ranges of the horizontal axis
are different for the 1995 data (left panels) and the 1996 data
(right panels.)}
\end{figure}

Several blue stars were discovered by KU and KR95 in NGC~6791, among
them B8. As was pointed out by KR95, the very blue color of B8 and its
variability strongly suggested a cataclysmic variable.  The photometry
of RKH supported this supposition. The star showed a distinct outburst
characteristic for dwarf novae stars which lasted a few days and had
an amplitude of about 2 mag in $V$ (Fig.1, left panel).  The star
became distinctly bluer during this outburst (see the CMD of the
cluster in Fig.5), which is typical for a stronger contribution
of the accretion disk to the combined color.  The data before the
outburst of B8 show relatively large scatter, much exceeding the random
errors of RKH photometry. They attempted a formal analysis for
presence of periodicities and saw a weak signal at 30.36 cycles per
day (47.4 minutes), with a small amplitude of 0.20 mag, only slightly
larger than the random errors of their photometry. This suspected
periodicity requires confirmation with more accurate data.

Additional $VI$ photometry for B8 was obtained by us during
September-October 1996 (Fig.1, right panel). The star showed overall
changes of luminosity between $V\sim 20.2$ and $V\sim 21.2$.  On some
nights we observed variability reaching $0.5\;mag$ on a time scale
shorter than $1\;hour$.

The spectrum of B8 (Fig.2) shows Balmer emission lines on a blue
continuum. A strong HeI line at 4471\AA\ is also seen, but no HeII
4686 is detected.  The bright Balmer lines are resolved and show
widths of about $2000\;km\;s^{-1}$ FWHM. The equivalent width of H$\beta$ is
30$\pm 5$\AA.  The $V$ magnitude estimated from the flux between 5000
and 6000\AA\ is 21.0, consistent with the variable being in its low
state. Overall, the spectrum is that of a cataclysmic variable in
quiescence.

\subsection{The cataclysmic variable B7}

\begin{figure}[t]
\plotfiddle{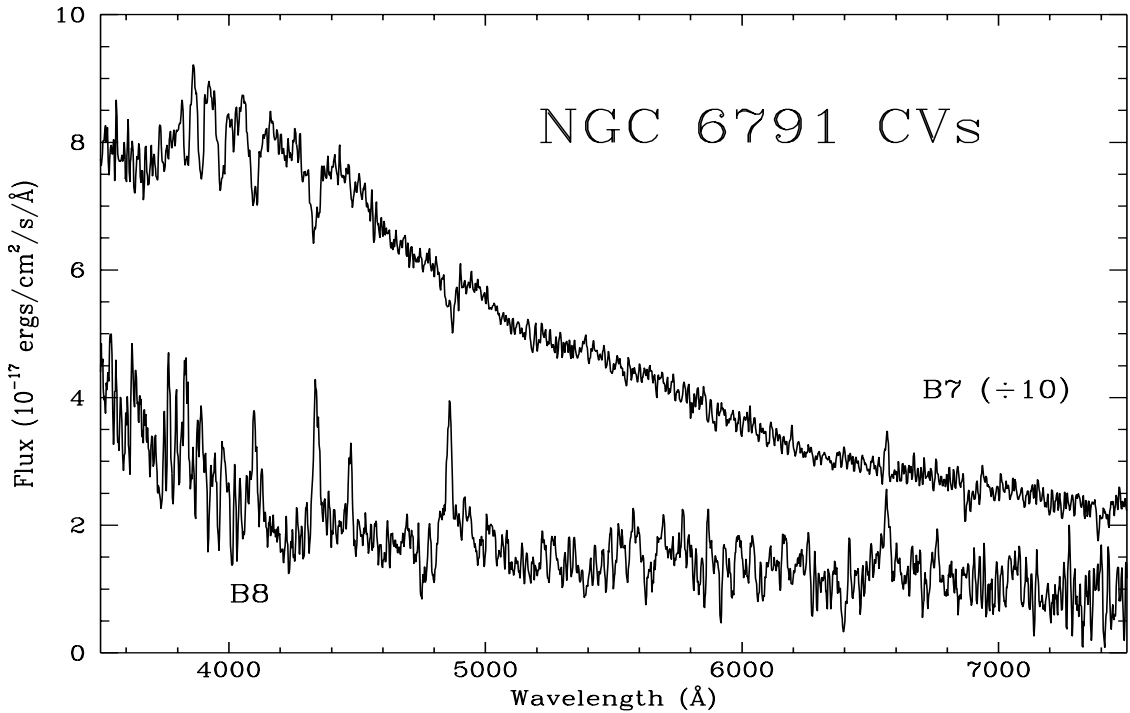}{8cm}{-90}{60}{50}{-240}{285}
\caption{MMT spectra of the variables B7 and B8 in NGC~6791. Please note 
that the flux of B7 was scaled down by a factor of 10.}
\end{figure}

\begin{figure}[p]
\plotfiddle{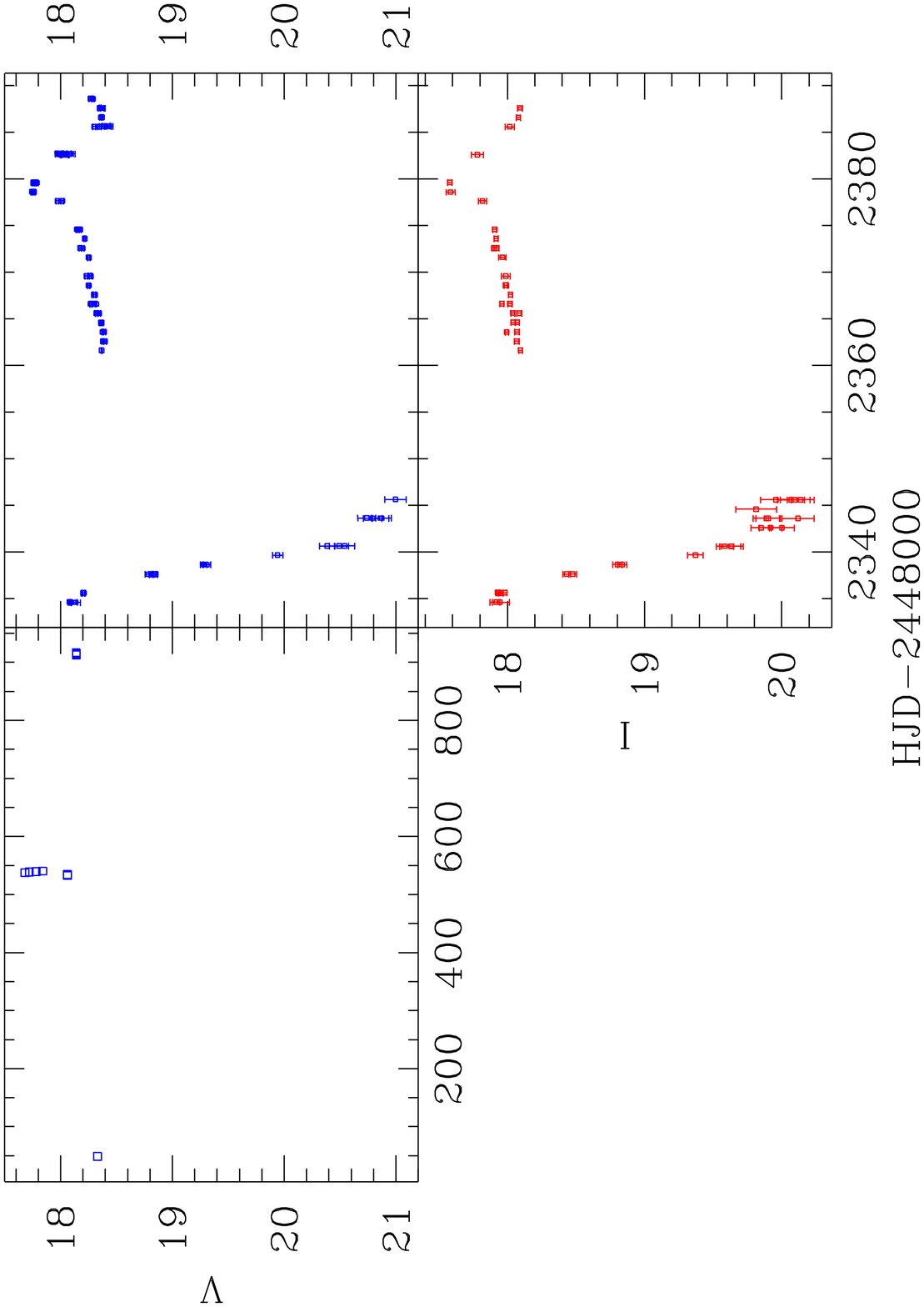}{8cm}{-90}{60}{50}{-240}{285}
\caption{$V$ (upper panels) and $I$ (lower panel) light curves of the
cataclysmic variable B7. Note that the ranges of the horizontal axis
are different for the 1990-92 data (left panel) and the 1996 data
(right panels.) }
\plotfiddle{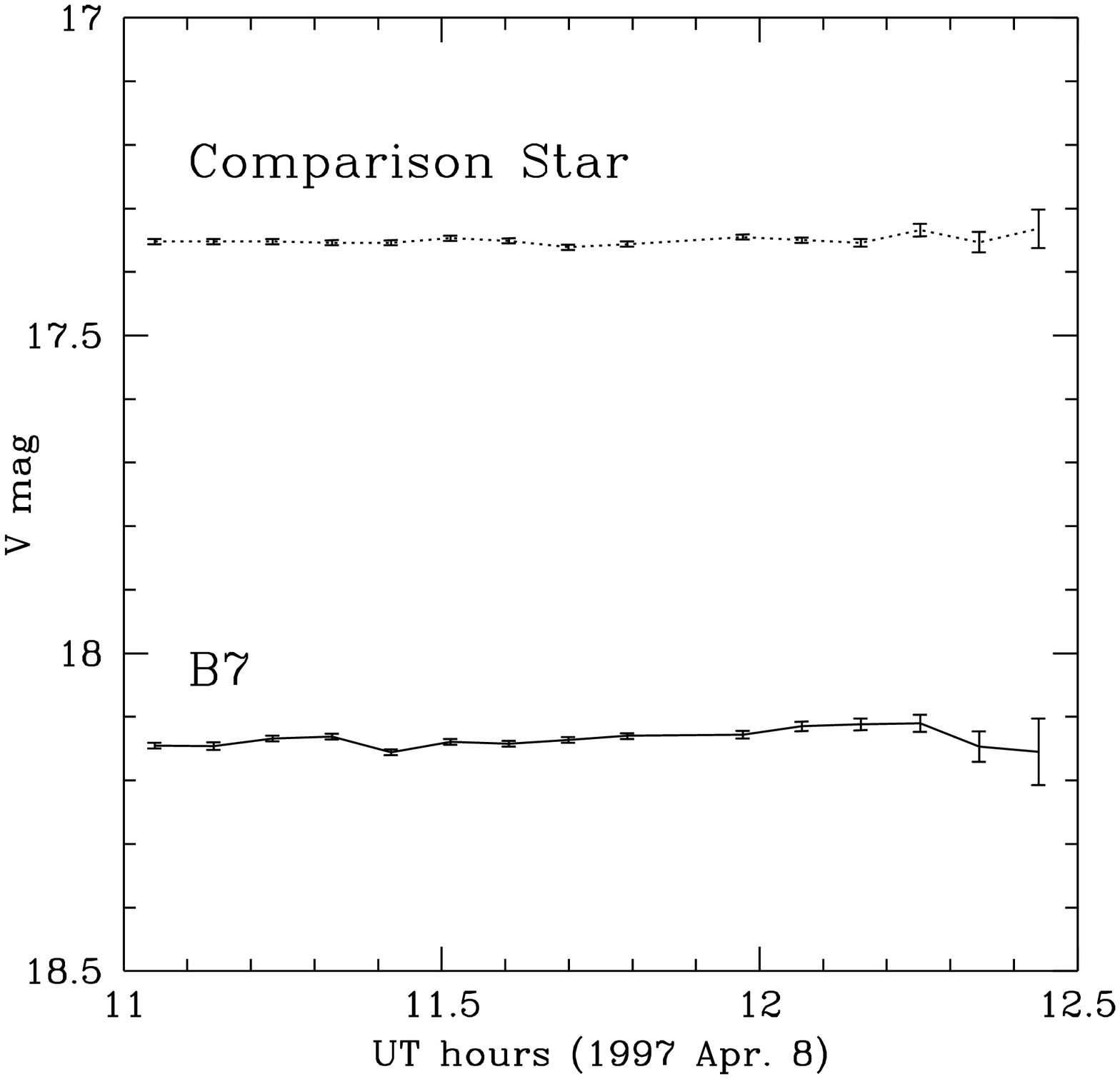}{9cm}{0}{50}{50}{-160}{-85}
\caption{$V$-band light curve of B7 obtained with the 2.1-m KPNO 
telescope on April 8, 1997.}
\end{figure}

B7 was also discovered by KU as one of the stars with very blue
color. KR93 found B7 to be weakly variable. The star changed its $V$
magnitude from 17.68 to 17.84 over 4 consecutive nights in October,
1991.  In the left panel of Fig.3 we show the light curve of B7 based
on the data published by KU, KR93 and KR95.  The stars was observed on a
total of 11 nights between June 1990 and October 1992.  B7 was not
observed by RKH as it fell outside their field of view.  Liebert et
al.~(1994) obtained spectroscopy for B7 and found it to have very
broad, shallow hydrogen and helium absorption lines, with narrow
emission cores, resembling a cataclysmic variable with an optically
thick accretion disk.

Our additional photometry obtained for B7 during September-October
1996 is presented in the right panel of Fig.3. The star underwent
a dramatic $>3\;mag$ drop in its brightness over the course
of $\sim10$ days and then went back to its ``high'' level
over the next $10-15$ days, although we did not observe
it on the rise. It then continued to vary slowly  throughout
October 1996, but with smaller variations of up to $0.6\;mag$.

Our spectrum of B7 taken on April 8, 1997, shows a blue continuum with
H$\beta$ and higher in absorption. The H$\alpha$ line is reversed and
seen as pure emission. The observed flux is consistent with the
spectrum being taken at optical maximum. The spectra taken in May show
only weak Balmer absorption and the flux implies that the CV was
caught outside a hight state with $V\approx 19$.

The variable B7 was also monitored for 1.5 hours on April 8, 1997,
using the KPNO 2.1-meter telescope and direct CCD camera. Fifteen
$300\;sec$ exposures were obtained with a standard Harris $V$ filter and
the results of the differential photometry are shown in Fig.4. The
dawn increase in sky brightness terminated the sequence,
and the effect of the rising background is reflected in the larger
error bars for the last few exposures.  The star showed very little
variability and there is no clear signature of flickering or orbital
modulation, although a longer monitoring interval is needed to rule
out the latter. The measured magnitude $V=18.13$ corresponds to B7
being in the high state, in agreement with the MMT spectrum taken on
the same night.

\section{DISCUSSION}

\begin{figure}[t]
\plotfiddle{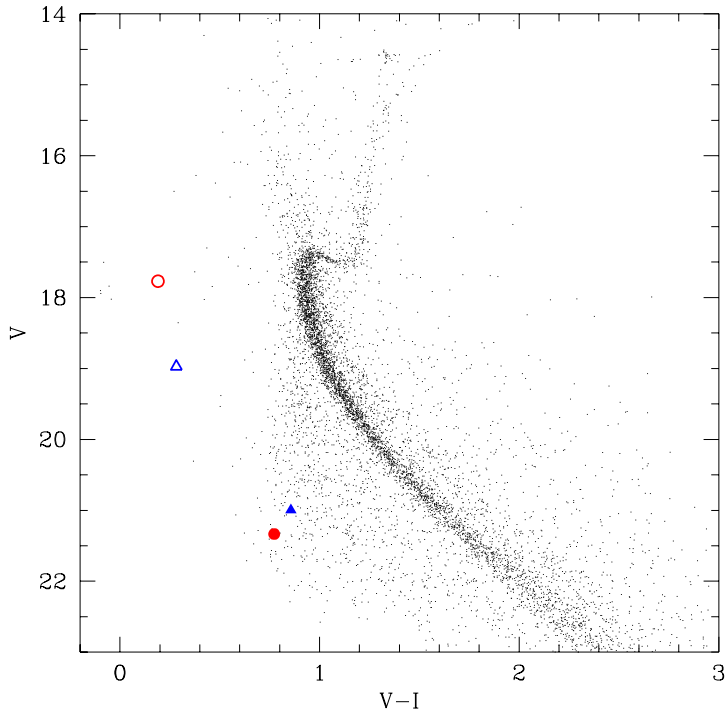}{8cm}{0}{50}{50}{-160}{-85}
\caption{$V,V-I$ color-magnitude diagram for the NGC~6791
cluster. Filled symbols represent the low states of B7 (triangle) and B8
(circle) variables and open symbols represent the high states
of these variables.}
\end{figure}

In Fig.5 we show a $V,V-I$ color-magnitude diagram for the NGC~6791
cluster, along with the low (filled symbols) and high (open symbols)
positions of B7 (triangle) and B8 (circle) variables.  The spectra of
B7 and B8 (Fig.2) demonstrate that these are cataclysmic variables.
The light curve of B8 suggests that it is a dwarf nova. The amplitude
for the outburst observed in 1995 was $2.0\;mag$ in $V$, which is a
relatively small range for dwarf novae. From the Kukarkin-Parenago
relationship, the outburst recurrence time for B8 should be in the
range of 10 to 30 days (Warner 1987). However, so far the variable was
caught only once in a clear outburst.

Photometrically B7 is seen in the high state most of the time, and
four spectra have failed to catch the CV at minimum. We speculate
that B7 is a nova-like CV which spend most of their time at a high mass
transfer rate with occasional dips to quiescence. Such behavior is
exhibited by some members of UX~UMa sub-type of CVs (Warner 1995).
The nova-like variables spending most of the time in a hight state, but
occasionally falling in brightness by one or more magnitudes are known
also as VY Scl stars. However, long-term monitoring is needed to
confirm this classification. Optionally, B7 can belong to Z Cam
subtype of CVs which have periods of ``standstill'' at magnitudes
intermediate between minimum and maximum.  The spectra of B7 obtained
at a hight state resemble spectra of some well studied UX~UMa type
stars. 

It is worth to note that B7, while in hight state, would be difficult
to identify as a CV using two methods most often applied to look for
the CVs in the centers of globular clusters. The first method is based
on searching for photometric variability. As can be seen in Fig.3, B7
was showing relatively little variability for 13 consecutive nights
after recovery from the low state.  A striking example of lack of
easily detectable variability in the light curve of B7 is show in
Fig.4.  The visual magnitude of B7 was constant to within about 0.035
during 1.5 hour of monitoring. The light curve is free of flickering,
a usual signature of light curves for most of the CVs.  The second
method often used to look for the CVs in clusters is based on
selecting objects with strong emission in Balmer lines.  Such objects
are selected based on a photometric index obtained from photometry
through broad band and narrow band filters.  One of our spectra of B7
(see Fig.2) shows only weak emission in $H\alpha$ and no emission at
all in higher lines of the Balmer series. The remaining spectra show
all of the Balmer lines in absorption.

The apparent distance modulus of NGC~6791 is $(m-M)_V=13.65$
(Garnavich et al.~1994) so we can convert the observed $V$-band
minimum and maximum into absolute magnitudes. Thus for B7 and B8 we
find minimum absolute magnitudes of $M_V=7.4$ and $M_V=7.6$,
respectively. The H$\beta$ equivalent width measured for B8 during
 quiescence predicts an absolute magnitude of $\sim 8$ from
Patterson's (1984) empirical fit, which is reassuringly close to our
measured value.

While we cannot determine the orbital periods of these two CVs from
their photometric variability, we can use their absolute brightnesses
at maxima to estimate periods based on the correlation given by Warner
(1987). For B7, the absolute visual magnitude at maximum is
$M_V=4.4\;mag$, so the period is expected to be approximately
$4.8\;hour$ for normal outbursts. If B7 were a Z~Cam star observed
during a standstill, then the orbital period would be substantially
longer. The absolute magnitude of the system at minimum places a
constraint on the maximum period if we assume the secondary fills its
Roche lobe. The secondary cannot be brighter than $M_V\sim 7.6\;mag$,
so the orbital period cannot be longer than $6\;hour$ if the secondary
is a normal dwarf (Caillault and Patterson 1990).

B8 is apparently fainter at maximum than B7, but only one maximum has
been observed and this makes the period estimate uncertain. For the
single maximum we find an absolute magnitude of $M_V\sim5.2$
corresponding to a period of $1.7\;hour$. This is below the period
gap and may mean that B8 is an SU~UMa star. Further monitoring is
needed to search for superoutbursts.

The two CVs discussed in this paper were originally selected based on
their blue colors and photometric variability.\footnote{
As may be seen in Fig.5, B7 and B8 exhibit relatively red V-I
colors while in low state.  B8 is, however, blue in $B-V$ and
particularly in $U-B$ color even when in low state (KR95). The same
holds most probably also in case of B7, although observational data to
support that are lacking at present.
}
None of the remaining nine blue stars with $V<19$ identified in
NGC~6791 by KU and KR95 had shown so far any evidence for variability
exceeding a few hundreds of magnitude. However, we note that several
faint blue objects with $V<21.5$ were detected in the cluster field by
KR95. If among these stars there were some intrinsically faint CVs they
would not be detected as variables neither in our survey nor in KR93
survey.  In particular, we note that AM~Her-type object detected in M67
by Gilliland et al.~(1991) would not have been detected by us as variable
star at the distance of NGC~6791.

It is remarkable that three CVs were discovered serendipitously in two
old open clusters (M67 and NGC~6791), while few CVs were
identified so far in dedicated surveys of centers of nearby globulars
performed with the HST (Grindlay et al.~1995). The samples of stars
observed in such clusters as NGC~6397 (Grindlay et al.~1995) and
NGC~6752 (Bailyn et al.~1996) are of comparable size to our sample of
NGC~6791 stars.  It is expected on theoretical grounds that large
numbers of CVs should be formed in GCs through two body tidal capture.
Central densities of NGC~6791 and particularly M67 are very low in
comparison with post-collapse clusters like NGC~6397 and
NGC~6752. Most likely CVs observed in M67 and NGC~6791 evolved from
primordial binaries and were not formed through two body tidal
capture.  Putting all this together, it seems that at least some
fraction of CVs observed in globulars also descend from primordial
binaries. That makes even more serious the discrepancy between
predictions of the tidal capture theory and observed low number of CVs
in GCs (eg.~Livio 1996).

We note parenthetically that NGC~6791 could be classified as a young,
low concentration globular cluster of extremely high metallicity.
Considering its richness, NGC~6791 is better populated than several
sparse globular clusters, such as E3 (Hesser et al.~1985). Moreover,
its luminosity function is flat for at least $5\;mag$ below the
turnoff (KR95), while the luminosity functions of other old open
clusters show turnover $3-4\;mag$ below the turnoff (Montgomery et
al. 1993; Caputo et al. 1990). Also the age of NGC~6791 is close to
ages of some ``young'' galactic globular clusters.  In particular the
principal sequences on the CMDs of NGC~6791 (KR95) and globular
cluster Ter~7 (Buonanno et al.~1995) match very well, if appropriate
shifts in color and magnitude are applied. The ages of these two
objects must be very close to each other. Buonanno et al.~(1994) noted
that four exceptionally young galactic globular clusters (Ter~7,
Arp~2, Ru~106 and Pal~12) are all located on a great circle on a
sky. They suggest that these young GCs could be captured by the Milky
Way from a companion galaxy.  NGC~6791 does not lay on the great
circle defined by positions of four clusters discussed by Buonanno et
al.~(1994).  However, the cluster does show a peculiar motion for an
open cluster.  It orbits the galactic center on highly eccentric orbit
with the galactocentric distance ranging from 4 to $10\;kpc$
(Anthony-Twarog 1996).  In respect to metallicity NGC~6791 resembles a
few super metal-rich globulars located close to the galactic center.
Taking into account high metallicity of NGC~6791 we may speculate that
it was formed near the galactic center as a massive cluster and then
kicked out on its present day orbit as a result of encounter with a
massive molecular cloud or interaction with a passing dwarf
galaxy. Such an event would lead to significant reduction of the mass
of the cluster.

\acknowledgments{We would like to thank the TACs of the 
MDM, the FLWO and the MMT Observatories for the generous amounts of
telescope time.  JK was supported by NSF grant AST-9528096 to Bohdan
Paczy\'nski and by the Polish KBN grant 2P03D011.12. KZS was supported
by the Harvard-Smithsonian Center for Astrophysics Fellowship.  We
thank Lisa Wells for the data taken with the KPNO 2.1-meter.}

\end{document}